\documentclass{amsart}

\newtheorem{theorem}{Theorem}[section]

\theoremstyle{definition}

\theoremstyle{remark}

\numberwithin{equation}{section}

\begin{document}



\title{A polynomial time algorithm for the Traveling Salesman problem}

\author{Sergey Gubin}






\begin{abstract}
The ATSP polytope can be expressed by asymmetric polynomial size linear program.
\end{abstract}

\maketitle

\section*{Introduction}
\label{s:intro}

In his seminal work \cite{yan}, M. Yannakakis proved that the Traveling Salesman problem (TSP) polytope cannot be expressed by symmetric polynomial size linear program, where symmetry means that the polytope is an invariant under vertex relabeling. The question about the size of asymmetric linear programs was left open in \cite{yan}, and it remained open since. 
\newline\indent
This article answers the question. We reduce the Asymmetric Traveling Salesman problem (ATSP) to  an asymmetric polynomial size linear program whose feasible set is asymmetric in the sense of \cite{yan}.

\section{Directed Hamiltonian Cycle Problem}
\label{s:dhc}

Directed Hamiltonian Cycle Problem (DHC) is a problem to find in any given digraph a cycle which visits all vertices (a circuit which visits all vertices and visits each of them just once). Any such cycle is called Hamiltonian.
\newline\indent
Ultimately, the problem can be solved by testing all vertex permutations on being a Hamiltonian cycle, or by ``growing'' of paths. The only drawback here is the number of ``elementary operations'' required in worst case\footnote{That number as a function of digraph's size is called a time or computational complexity.}. Computational complexity of the methods will be factorial and exponential appropriately. So, they are infeasible for modern computers even in the cases of modest digraphs. 
\newline\indent
The real problem is to detect Hamiltonian cycles in feasible time. Sometimes, that can be done using the appropriate criterion \cite[and many others]{Hamilton, Tutte1, Ore, Chvatal, Bollabas, Fan, Bauer2}. But, such particular instances of DHC are rare.
\newline\indent
The importance of DHC follows from the fact that its decision version\footnote{The problem of the existence of Hamiltonian cycles.} is a NP-complete problem\footnote{NP-complete problem is such NP-problem to which any other NP-problem can be reduced in polynomial time by deterministic Turing machine (DTM). DTM is a rigorously defined algorithm. NP-problem is a problem which can be solved in polynomial time by non-deterministic Turing machine (NDTM). NDTM is an aggregate of a non-deterministic polynomial time generator of guesses for solution and a polynomial time DTM which checks those guesses on being a solution. P-problem is a problem which can be solved by DTM in polynomial time. For more information, see, for example, \cite{sf}.} \cite{karp}. There is a wide spread belief in P $\neq$ NP. This hypothesis means that no polynomial time algorithm exists for DHC \cite{karp, sf}. Yet, below we express DHC by a polynomial size linear system. It is well known that solution of such systems is a P-problem \cite{khach}.

\subsection{DHC as a relabeling problem}

Let $G$ be a given digraph with $n$ vertices, $n > 1$. Let's arbitrarily enumerate/label vertices of $G$. Let $A_G$ be the appropriate adjacency matrix of $G$:
\[
A_G = (g_{ij})_{n\times n}
\]
- where $g_{ij}$ is $0$ or $1$ depending on the existence/absence of arc from vertex $i$ into vertex $j$.
\newline\indent
Obviously, there is a Hamiltonian cycle in digraph $G$ iff there is a circular permutation submatrix in matrix $A_G$. That submatrix is an adjacency matrix of that cycle. This trivial fact may be expressed by the following quadratic system\footnote{For two matrices $B=(b_{ij})$ and $C=(c_{ij})$ of the same size, relation $B \geq C$ means that $\forall i,j ~ (b_{ij} \geq c_{ij})$.}:
\begin{equation}
\label{e:quadro}
A_G \geq X S X^T,
\end{equation}
- where unknown $X$ is a permutation matrix, and constant $S$ is any circular permutation matrix - an adjacency matrix of a Hamiltonian cycle. Let's fix to $S$ the following value:
\[
S = (s_{ij})_{n\times n} = \left ( \begin{array}{ccccc}
0 & 1 & 0 & \ldots & 0 \\
0 & 0 & 1 & \ddots & 0\\
\vdots & \vdots & \ddots & \ddots & \vdots \\
0 & 0 & 0 & \ldots & 1\\
1 & 0 & 0 & \ldots & 0\\
\end{array} \right )_{n\times n}
\]
\indent
Any permutation-matrix solution $X$ of system \ref{e:quadro} means a Hamiltonian cycle in $G$. Matrix $XSX^T$ is an adjacency matrix of that cycle, and matrix $X^T$ defines relabeling of $G$ along it. After that relabeling, this cycle will visit vertices of $G$ in the following order: 
\[
1 \rightarrow 2 \rightarrow 3 \rightarrow \ldots \rightarrow n \rightarrow 1
\]
\indent
System \ref{e:quadro} presents DHC in all senses: DHC as a decision problem is the consistence problem for the system; DHC as a search problem\footnote{The problem of the actual finding of Hamiltonian cycles.} is the problem to solve the system; DHC as a counting problem\footnote{The problem to count different Hamiltonian cycles in $G$ - $n$ times of that number is the number of solutions of system \ref{e:quadro}.} is the problem to count the system's solutions, etc.
\newline\indent
In terms of system \ref{e:quadro}, guesses are vertex permutations.  They are presented in the system with unknown permutation matrix $X$. Each solution of the system delivers a Hamiltonian cycle.

\subsection{Compatibility matrix}

In system \ref{e:quadro}, digraph $G$ is presented with its adjacency matrix. Adjacency matrix is a universal encoding for digraphs. To solve system \ref{e:quadro}, we will use special encoding which we call a \emph{compatibility matrix}.
\newline\indent
For each two vertices $i$ and $j$, let's build a \emph{compatibility box}. The box is the following matrix\footnote{Here, we intentionally disregard the type of $S$ and that particular value we fixed to it.} $C_{ij} = (c_{ij\mu\nu})_{n\times n}$:
\begin{equation}
\label{e:box}
c_{ij\mu\nu} = \left \{ \begin{array}{cl}
1, & s_{ij} \leq g_{\mu\nu} \wedge c_{ji} \leq g_{\nu\mu} \\
0, & s_{ij} > g_{\mu\nu} \vee s_{ji} > g_{\nu\mu} \\
\end{array} \right.
\end{equation}
The $(\mu,\nu)$-th element of $C_{ij}$ indicates whether vertex couple $(\mu,\nu)$ can be relabeled into couple $(i,j)$ regardless of anything but adjacency. 
\newline\indent
There are $n^2$ compatibility boxes, and it takes time $O(n^2)$ to compute any of them with brute force. The boxes have the following major properties: 
\begin{equation}
\label{e:boxproperty}
\begin{array} {c}
C_{ji} = C_{ij}^T \\
C_{ii} = U_n \\
i \neq j \Rightarrow c_{ij\mu\mu} = 0 \\
\end{array}
\end{equation}
- where $U_n$ is $n\times n$ identity matrix, and all indexes are in their ranges.
\newline\indent
Compatibility matrix is a box matrix which aggregates all compatibility boxes in accordance with their indexes:
\[
C = (C_{ij})_{n\times n}
\]
In elements, the matrix has size $n^2 \times n^2$.
\newline\indent
Compatibility matrix aggregates all relabeling options for vertex couples of $G$. For our particular value of $S$, the matrix looks:
\[
C = \left ( \begin{array}{cccccc}
U_n & A_G & K_n  & \ldots & K_n & A_G^T \\
A_G^T & U_n  & A_G& \ddots & K_n & K_n \\
K_n  & A_G^T & U_n  & \ddots & K_n & K_n \\
\vdots & \ddots & \ddots & \ddots & \ddots & \vdots \\
K_n  & K_n & K_n  & \ddots & U_n & A_G \\
A_G & K_n  & K_n & \ldots & A_G^T & U_n \\
\end{array} \right )_{n\times n}
\]
- where the size of $C$ is shown in boxes and 
\[
K_n = (1)_{n\times n} - U_n
\]
Obviously, the pattern holds for any circular permutation matrix $S$.
\newline\indent
Due to the previous subsection, $G$ has a Hamiltonian cycle iff its vertex couples can be relabeled without any contradiction, i.e. iff $C$ has a grid of non-zero elements, one element per compatibility box. Any such grid of elements in the compatibility matrix we call a \emph{solution grid}. The compatibility matrix encoding of $G$ reduces DHC to search the encoding for solution grids.
\newline\indent
All elements of any solution grid equal $1$, their indexes $\mu$ depend on their indexes $i$ and their indexes $\nu$ depend on their indexes $j$, only\footnote{Index $\mu$ is a function of index $i$, and index $\nu$ is a function of index $j$.}:
\begin{equation}
\label{e:grid}
\{c_{ij\mu\nu} = 1 ~|~ \mu = \mu(i), ~\nu = \nu(j), ~ i,j=1,2,\ldots,n \}
\end{equation}
The following major properties of the elements' indexes follow from properties \ref{e:boxproperty} of the compatibility boxes:
\begin{equation}
\label{e:gridproperty}
\begin{array}{ccc}
i = j & \Rightarrow & \mu(i) = \nu(j) \\
i_1 \neq i_2 & \Rightarrow & \mu(i_1) \neq \mu(i_2) \\
j_1 \neq j_2 & \Rightarrow & \nu(j_1) \neq \nu(j_2) \\
\end{array}
\end{equation}
- where all indexes are in their ranges. Due to these properties, any solution grid is a cyclical permutation of vertex indexes:
\[
(\mu(1) : \mu(2) : \mu(3) : \ldots : \mu(n))
= 
(\nu(1) : \nu(2) : \nu(3) : \ldots : \nu(n))
\]
In other words, it is a Hamiltonian cycle in $G$:
\[
\mu(1) ~ \rightarrow ~ \mu(2) ~ \rightarrow ~ \mu(3) ~ \rightarrow ~   \ldots ~ \rightarrow ~ \mu(n) ~\rightarrow ~\mu(1)
\]
\indent
Now, let's describe the guessing in the compatibility matrix's terms. It may be organized as follows:
\newline
(1) Guess $\gamma$ is a box matrix with the same structure as $C$. In the matrix, all elements equal $0$ except those which create a possible solution grid in $C$, i.e. except those whose indexes create a set 
\[
\{(i,j,\mu,\nu) ~|~ \mu = \mu(i), ~\nu = \nu(j), ~ i,j=1,2,\ldots,n \}
\]
- where functions $\mu(i)$ and $\nu(i)$ satisfy properties \ref{e:gridproperty}. All elements with these indexes equal $1$. 
\newline
(2) There are $n!$ such guesses $\gamma$ in total. Let $\Gamma$ be set of all of them:
\begin{equation}
\label{e:gamma}
\Gamma = \{ \gamma_1, \gamma_2, \ldots, \gamma_{n!}\}
\end{equation}
(3) Checking conditions are the following relation:
\begin{equation}
\label{e:test}
\gamma \leq C, ~ \gamma \in \Gamma
\end{equation}
Non-zero elements of any guess $\gamma$ which passed this test are a solution grid in $C$.

\subsection{Convex hull of all guesses}

The following theorem clarifies the benefits of the compatibility matrix encoding.
\begin{theorem}
\label{t:hull}
Convex hull of guesses \ref{e:gamma} is the following polytope:
\begin{equation}
\label{e:hull}
\left \{ \begin{array}{l}
x_{ij\mu\nu} = x_{ji\nu\mu},~ x_{ij\mu\nu} \geq 0 \\
\mbox{- where}~ i,j,\mu,\nu = 1,2,\ldots,n, ~ i \neq j, ~ \mu \neq \nu \\
\\
\sum_{\mu=1,~\mu\neq\nu}^n x_{ij\mu\nu} = y_{jj\nu\nu} \\
\mbox{- where} ~ i,j,\nu = 1,2,\ldots,n, ~ i \neq j\\
\\
\sum_{i=1, ~ i\neq j}^n x_{ij\mu\nu} = y_{jj\nu\nu} \\
\mbox{- where} ~ j,\mu,\nu = 1,2,\ldots,n, ~ \mu \neq \nu \\
\\
\sum_{\nu=1}^n y_{jj\nu\nu} = 1, ~ y_{jj\nu\nu} \geq 0 \\
\mbox{- where} ~ j = 1,2,\ldots, n
\end{array} \right.
\end{equation}
\end{theorem}
\begin{proof}
System \ref{e:hull} may be described with the following box matrix of unknowns:
\[
B = \left ( \begin{array} {cccc}
Y_{1,1} & X_{1,2} & \ldots & X_{1,n} \\
X_{2,1} & Y_{2,2} & \ddots & X_{2,n} \\
\vdots & \ddots & \ddots & \vdots \\
X_{n,1} & X_{n,2} & \ldots & Y_{n,n} \\
\end{array} \right )_{n^2 \times n^2}
\]
Diagonal boxes in $B$ are the following $n\times n$ diagonal matrices:
\[
Y_{ii} = \mbox{diag}(y_{i,i,1,1} ~ y_{i,i,2,2} ~ \ldots ~ y_{ii\nu\nu} ~ \ldots ~ y_{i,i,n,n})
\]
Off-diagonal boxes in $B$ are the following $n\times n$ matrices :
\[
X_{ij} = \left ( \begin{array}{cccc}
0 & x_{i,j,1,2} & \ldots & x_{i,j,1,n} \\
x_{i,j,2,1} & 0 & \ddots & x_{i,j,2,n} \\
\vdots & \ddots & \ddots & \vdots \\
x_{i,j,n,1} & x_{i,j,n,2} & \ldots & 0 \\
\end{array} \right )_{n\times n}
\]
System \ref{e:hull} states the following relations between elements of $n^2\times n^2$ matrix $B$:
\newline
(1) $B$ is a symmetric matrix: $X_{ji} = X_{ij}^T$. Thus, all the below will be true in the horizontal direction as well as in the vertical direction;
\newline
(2) For every fixed box column $j$ and for every fixed column $\nu$ in it: the total over $\mu = 1,2,\ldots, n$, $\mu \neq \nu$, of all elements $x_{ij\mu\nu}$ does not depend on $i$. It equals to element $y_{jj\nu\nu}$ of box $Y_{jj}$;
\newline
(3) For every fixed box column $j$, for every fixed column $\nu$ in it, and for every fixed row $\mu \neq \nu$, the total over $i \neq j$ of all elements $x_{ij\mu\nu}$ does not depend on $\mu$. It equals to element $y_{jj\nu\nu}$ of box $Y_{jj}$;
\newline
(4) For every box $Y_{jj}$, its diagonal is a convex decomposition of $1$.
\newline\indent
Let's notice, system \ref{e:hull} is consistent. For example, the following solution of the system we call a \emph{center}:
\[
y_{jj\nu\nu} \equiv \frac{1}{n}, ~ x_{ij\mu\nu} \equiv \frac{1}{n(n-1)}
\]
- where all indexes are in their ranges. The center minimizes Euclidean norm in polytope \ref{e:hull}. Other obvious solutions of system \ref{e:hull} are guesses \ref{e:gamma}. The guesses maximize Euclidean norm in polytope \ref{e:hull}. Thus, every guess is a vertex of the polytope.
\newline\indent
Now, let the above matrix $B$ be a solution of system \ref{e:hull}. It is easy to see that, for every fixed $j$ and $\nu$, elements $x_{ij\mu\nu}$, $i\neq j$ and $\mu\neq\nu$, constitute a $(n-1)\times (n-1)$ doubly stochastic matrix multiplied on $y_{jj\nu\nu}$ (the same is true for elements $x_{ji\nu\mu}$, $i\neq j$ and $\mu\neq\nu$). Due to Birkhoff-von Neumann theorem \cite{bir}, that matrix is a convex combination of $(n-1)\times (n-1)$ permutation matrices multiplied on $y_{jj\nu\nu}$. Let's substitute these convex decompositions in $B$ and write $B$ as the appropriate sum. That sum is a convex decomposition of $B$ over guesses \ref{e:gamma}. Thus, the guesses are the only vertices in polytope \ref{e:hull}.
\end{proof}

\subsection{The ATSP polytope}

Due to theorem \ref{t:hull}, solution grids are those vertices of polytope \ref{e:hull} which satisfy inequality \ref{e:test}. But, that inequality can be expressed with the following equalities:
\begin{equation}
\label{e:atsp}
x_{i_0 j_0 \mu_0 \nu_0} = 0
\end{equation}
- where indexes $(i_0, j_0, \mu_0, \nu_0)$ are indexes of all those elements in compatibility matrix $C$ which equal $0$.
\begin{theorem}
\label{t:dhc}
Convex hull of all solution grids is polytope \ref{e:hull}, \ref{e:atsp}.
\end{theorem}
\begin{proof}
Let's just notice that equations \ref{e:atsp} cut from polytope \ref{e:hull} a chunk. That cutting is along the $(i_0, j_0, \mu_0, \nu_0)$-th coordinate lines going through vertex points of the polytope, i.e. it does not create new vertices.
\end{proof}
Aggregated linear system \ref{e:hull} and \ref{e:atsp} expresses the ATSP polytope. The system has polynomial size and can be solved in polynomial time. For example, Khachiyan's ellipsoid algorithm \cite{khach, gls} and Karmarkar's interior-point algorithm \cite{karmarkar} will solve this system in strongly polynomial time, because all its coefficients are $0$ or $1$.
\newline\indent
Due to definition  \ref{e:box} of the compatibility boxes, system \ref{e:atsp} explicitly involves an adjacency matrix of digraph $G$. Vertex relabeling of $G$ will rotate chunk \ref{e:atsp} all over polytope \ref{e:hull}. In other words, aggregated system \ref{e:hull} and \ref{e:atsp} is asymmetric in the sense of \cite{yan}. Thus, theorem \ref{t:dhc} may be seen as complimentary to Yannakakis' theorem \cite{yan}: the ATSP polytope can be expressed by an asymmetric polynomial size linear program.
\newline\indent
Let's notice that, due to theorem \ref{t:hull}, any solution of aggregated system \ref{e:hull} and \ref{e:atsp} can be presented as a convex combination of guesses \ref{e:gamma}. This decomposition is a P-problem. Due to equalities \ref{e:atsp}, any guess in any such decomposition will be a solution grid, i.e. it will deliver a Hamiltonian cycle. Thus, theorem \ref{t:dhc} efficiently solves DHC in both senses as a decision problem and as a search problem.
\newline\indent
Also, let's notice that the number of vertices in polytope \ref{e:hull}, \ref{e:atsp} divided by $n$ is the number of different Hamiltonian cycles in $G$.

\section{Asymmetric Traveling Salesman Problem}

Asymmetric Traveling Salesman Problem (ATSP) is a problem to find in any given weighted digraph a Hamiltonian cycle with minimal total weight.
\newline\indent
ATSP is a well known problem of combinatorial optimization \cite[and many others]{yan, dantzig, arora, tsp, wcook, soda}. Decision version\footnote{The problem of existence of a Hamiltonian cycle with total weight in any given boundaries.} of ATSP is a NP-complete problem \cite{karp}. Yet, theorem \ref{t:dhc} allows an expression of ATSP by a polynomial size linear program.
\newline\indent
Let $W$ be a weight function on given digraph $G$:
\[
W = (w_{ij})_{n\times n}: (i,j) ~ \mapsto ~ w_{ij} \in (-\infty,+\infty], ~i,j = 1,2,\ldots,n
\]
- where $w_{ij}$ is the weight of arc from vertex $i$ into vertex $j$ (as usual, $w_{ij} = +\infty$ if there is not any such arc). 
\begin{theorem}
\label{t:atsp}
The minimal total weight of Hamiltonian cycles in $G$ is the solution of the following linear program:
\begin{equation}
\label{e:min}
\sum_{i,j,\mu,\nu = 1}^n w_{ij}x_{ij\mu\nu} ~ \rightarrow ~ \min
\end{equation}
- subject to constrains \ref{e:hull} and \ref{e:atsp}.
\end{theorem}
\begin{proof}
Due to theorem \ref{t:dhc}, $G$ has Hamiltonian cycles iff program \ref{e:min} has non-empty feasible set. Due to the same theorem, vertices of that set are Hamiltonian cycles in $G$, and the addends in criterion \ref{e:min} pertain to the weights of those arcs which participate in the cycles.
\end{proof}
Linear program \ref{e:min} has polynomial size and can be solved in polynomial time \cite{khach, gls, karmarkar}. 
From the practical perspective, let's notice that we do not require weights $w_{ij}$ to be positive.

\section*{Conclusion}

In this article, we reduced DHC to an asymmetric $O(n^4)$-size linear system and expressed ATSP by the appropriate linear program.
\newline\indent
The linearization was done by immersing $O(n^2)$-dimensional algebraic variety \ref{e:quadro} in $R^{n^4}$. The immersion was done with the compatibility matrix encoding. 
\newline\indent
Compatibility matrix is an encoding of DHC instances in the contradictions between relabeling options for vertex couples. The options can be computed in $O(n^4)$-time with the brute force method. Analysis of the contradictions is a parallel testing of all guesses. The testing is the solution of linear system \ref{e:hull} and \ref{e:atsp}.
\newline\indent
Let's notice that the role of those particular value and type of matrix $S$ in inequality \ref{e:quadro} was insignificant. So, our method can be directly applied to Subgraph Isomorphism problem (SubGI) in (multi) digraphs with (multi) loops \cite{subgi} and to related optimizations.
\newline\indent
Because of the possibility of loops in SubGI (the diagonal elements in adjacency matrices $A_G$ and $S$ may be positive), the second property \ref{e:boxproperty} has to be changed to inequality
\[
C_{ii} \leq U_n
\]
Strong inequalities in these relations cause additional constrains on variables $y_{ii\mu\mu}$ in system \ref{e:atsp}:
\[
y_{i_0 i_0 \mu_0 \mu_0} = 0
\]
That will be the only change for SubGI.
\newline\indent
There is a demo \cite{demo}. Using it, readers may try the reduction on their own examples.

\end{document}